# Observation of new magnetic ground state in frustrated quantum antiferromagnet spin liquid system Cs$_2$CuCl$_4$

Hyeong-Jin Kim[1,2*], C. R. S. Haines[1*], C. Liu[1], Sae Hwan Chun[2], Kee Hoon Kim[2], H. T. Yi[3], Sang-Wook Cheong[3] & Siddharth S. Saxena[1*]

[1]*Quantum Matter group, Cavendish Laboratory, University of Cambridge, Madingley Road, Cambridge CB3 0HE, United Kingdom.*

[2]*CeNSCMR, Department of Physics and Astronomy, Seoul National University, Seoul 151-747, Republic of Korea.*

[3]*Department of Physics & Astronomy, Rutgers University, Piscataway, New Jersey 08854, USA.*

**Cs$_2$CuCl$_4$ is known to possess a quantum spin liquid phase with antiferromagnetic interaction below 2.8 K. We report the observation of a new metastable magnetic phase of the triangular frustrated quantum spin system Cs$_2$CuCl$_4$ induced by the application of hydrostatic pressure. We measured the magnetic properties of Cs$_2$CuCl$_4$ following the application and release of pressure after 3 days. We observed a previously unknown ordered magnetic phase with a transition temperature of 9 K. Furthermore, the recovered sample with new magnetic ground state possesses an equivalent crystal structure to the uncompressed one with antiferromagnetic quantum spin liquid phase.**

Geometrically frustrated spin structures have been the focus of great attention from a broad range of the scientific community due to the potential functional and fundamentally interesting physical properties they display such as superconductivity, magnetism and ferroelectricity [1–6]. The physical properties result from the competition between exchange coupling strengths on the frustrated antiferromagnetic spin lattice. The magnetic phases and phase transitions of frustrated antiferromagnetic quantum spin systems have been investigated experimentally and theoretically [1–8].

In this article, we report the discovery of a new magnetic ground state in the frustrated quantum antiferromagnet spin-1/2 system, Cs$_2$CuCl$_4$, obtained by holding the sample under hydrostatic pressure of approximately 5 kbar at room temperature for 2 days or more. The new phase displays magnetic properties that are similar to chiral helimagnet-like behaviour in stark contrast to the quantum spin liquid phase of unpressurised Cs$_2$CuCl$_4$. Furthermore, there is no significant structural distortion or phase transition accompanying this new magnetic structure.

Cs$_2$CuCl$_4$ is a well-studied realization of a two-dimensional quantum spin-liquid phase [10]. The crystal structure of Cs$_2$CuCl$_4$ is orthorhombic with space group *Pnma* [15]. It is composed of CuCl$_4^{2-}$ tetrahedra and Cs$^+$ ions as shown in figure 1(a). It provides a spatially anisotropic spin-1/2 triangular antiferromagnet, which is characterized by Cu$^{2+}$ ions on a geometrical *bc* plane. The interchain exchange coupling $J'$ (= 0.125 meV) between Cu$^{2+}$ ions along zigzag directions on the triangular *bc* plane is weaker than the intrachain exchange coupling $J$ (= 0.375 meV) along *b* axis, with the coupling ratio of $J'/J = 0.34$ [10, 16, 17]. Experimentally, the magnetic property of Cs$_2$CuCl$_4$ has been assigned to be a two-dimensional quantum spin-liquid phase, by neutron scattering [16 - 18], magnetization [19] and specific heat [20] measurements.

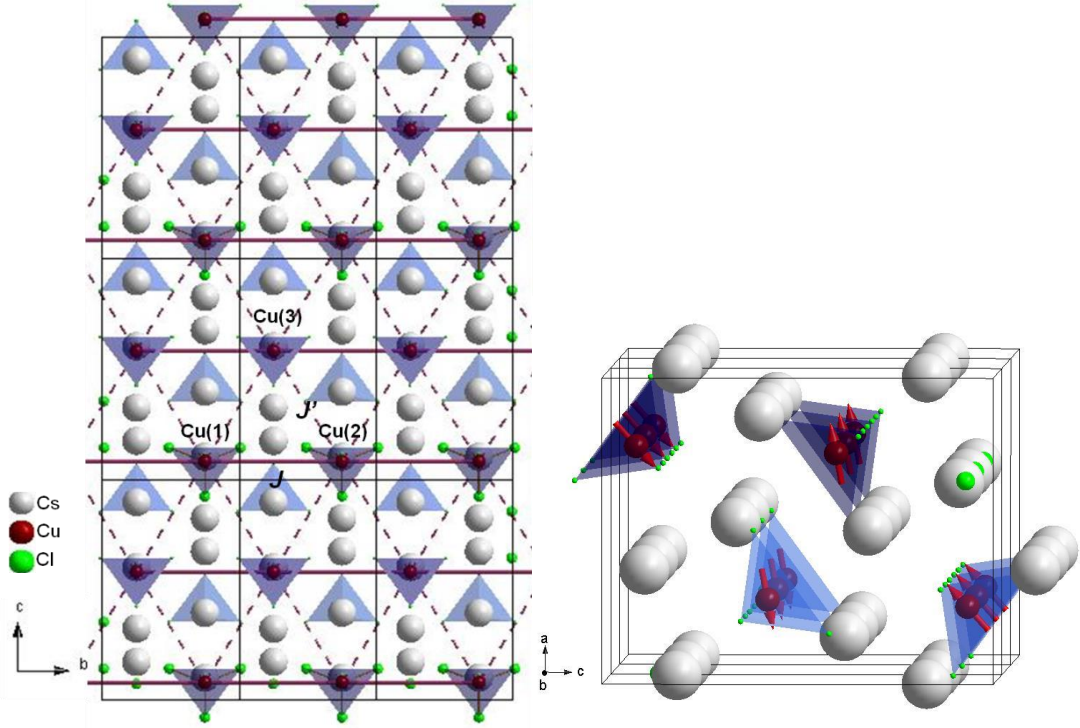

**Figure 1: Crystal structure of $Cs_2CuCl_4$.** (a) Orthorhombic crystal structure for the frustrated quantum antiferromagnet spin-1/2 $Cs_2CuCl_4$ with $Cu^{2+}$ surrounded by $Cl^-$ ions and $Cs^+$. The two antiferromagnetic spin interactions are the nearest neighbour exchange coupling $J$ between Cu(1) and Cu(2) in the b-direction and the next nearest neighbour exchange coupling $J'$ between Cu(1) and Cu(3) on the triangular $bc$ plane with a ratio of $J'/J = 0.34$. (b) spin orientation (grey arrows) in Cu atoms on the $ac$ plane.

The DC magnetic susceptibility of pre-pressurized and post- pressurised $Cs_2CuCl_4$ is shown in Fig. 2. In the pre-pressurised sample the well-known phase transition from paramagnet to two-dimensional quantum spin-liquid phase with short-range antiferromagnetic order around $T_m = 2.8$ K is observed. We pressurized the sample to 5 kbar for 4 days in a hydrostatic piston cylinder cell and then released the pressure (post-pressurized $Cs_2CuCl_4$). The susceptibility behaviour of the post-pressurized $Cs_2CuCl_4$ is significantly different from that of the pre-pressurized sample. The susceptibility is enhanced by nearly an order of magnitude and a new transition is observed at a temperature of $T_s = 9$ K. The post-pressurized phase has a very sharp increase with decreasing temperature to a large maximum moment around $T_s = 9$ K. From fitting the high temperature (T ≥20 K) behaviour using a Curie-Weiss formula $\chi = C/(T+\Theta)$, a Curie temperature of $\Theta = 3.3 \pm 0.6$ K for pre-pressurized $Cs_2CuCl_4$ was obtained, consistent with the reported value [27], while that of $\Theta = -7.1 \pm 0.3$ K was obtained for the post-pressurized phase. This implies that the spin exchange interaction in $Cs_2CuCl_4$ changes from antiferromagnetic to ferromagnetic-type after pressurisation. As shown in the upper inset of figure 2, the susceptibility of the post-pressurized $Cs_2CuCl_4$ exhibits reversible behaviour as field-cooled and zero-field-cooled data are indistinguishable.

In order to investigated the reproducibility and pressurisation time dependence of the new magnetic state in $Cs_2CuCl_4$, samples of $Cs_2CuCl_4$ were pressurized to 5kbar for increasing amounts of time ranging from 0 to 6 days. The new phase was observed in the samples pressed under about 5 kbar and released after more than two days, as shown in figure 2(b) (solid symbol). Also, in order to verify the stability of the new magnetic phase for the post-pressurized

Cs$_2$CuCl$_4$, we measured its susceptibility after keeping it in an evacuated desiccator for more than one month. We found no change in magnetic properties after this period. This demonstrates that this new magnetic state can be stabilized after a sluggish transition period and is a new magnetic ground state in the triangular frustrated antiferromagnet spin-1/2 system.

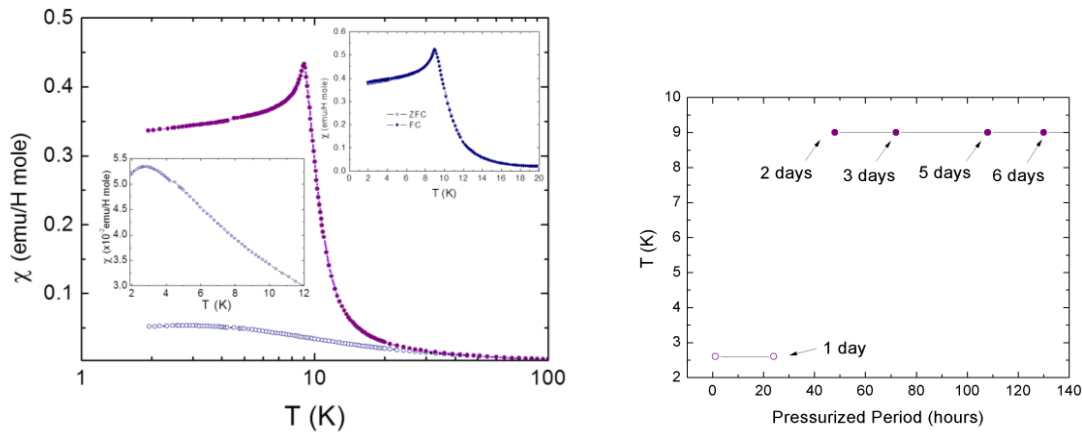

**Figure 2: Susceptibility of pre- and post-pressurized Cs$_2$CuCl$_4$ with pressing time dependence.** (a) The dc susceptibility for pre-pressurized (open symbol) and post-pressurized (solid symbol) Cs$_2$CuCl$_4$ with external magnetic field parallel to ac plane under ambient pressure. The susceptibility of the post-pressurized sample shows a transition temperature of $T_s = 9$ K. The upper inset shows that there is no difference between the zero-field-cool and field-cool susceptibilities. (b) The new ground magnetic phase of post-pressurized Cs$_2$CuCl$_4$ could be observed under ambient pressure when it was pressed for more than 2 days.

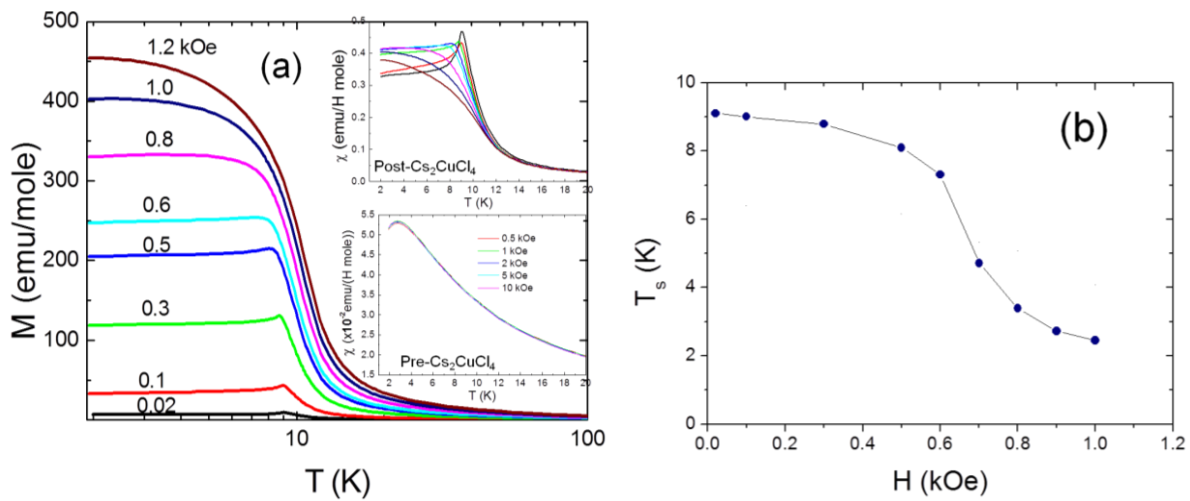

**Figure 3: Magnetization of the post-pressurized Cs$_2$CuCl$_4$.** Temperature dependence of magnetisation curves of the post-pressurized Cs$_2$CuCl$_4$ with each field (parallel to ac plane) from 0.02 to 1.2 kOe under ambient pressure in the main figure (a). The transition temperature is about 9 K up to 0.1 kOe. But as field increase, the peak transition temperature, $T_p$, is decreased and, above 1 kOe, has disappeared (b). The upper inset in figure (a) shows the susceptibility from the magnetisation curve normalized by each field for the post-pressurized Cs$_2$CuCl$_4$. The lower inset in figure (a) shows the susceptibility of the pre-pressurized Cs$_2$CuCl$_4$ without change of transition temperature for several magnetic fields.

Figure 3(a) shows the dc magnetisation curves of post-pressurized $Cs_2CuCl_4$ as a function of temperature at different magnetic fields parallel to the *ac* plane. As the applied magnetic field is increased up to 1 kOe, the transition temperature $T_s$ is seen to decrease and broaden, and could not be observed above 2 K at above $H = 1$ kOe as shown in figure 3(b). The magnetization curves show a saturated behaviour above 1 kOe at low temperature, which may suggest a field-induced phase transition to a ferromagnetic order phase. As shown in the lower inset of figure 3(a), these field-dependent susceptibilities, determined from magnetisation data, are also quite different from those for the pre-pressurized $Cs_2CuCl_4$ which exhibited field-independent susceptibility as shown in the upper inset of figure 3(a).

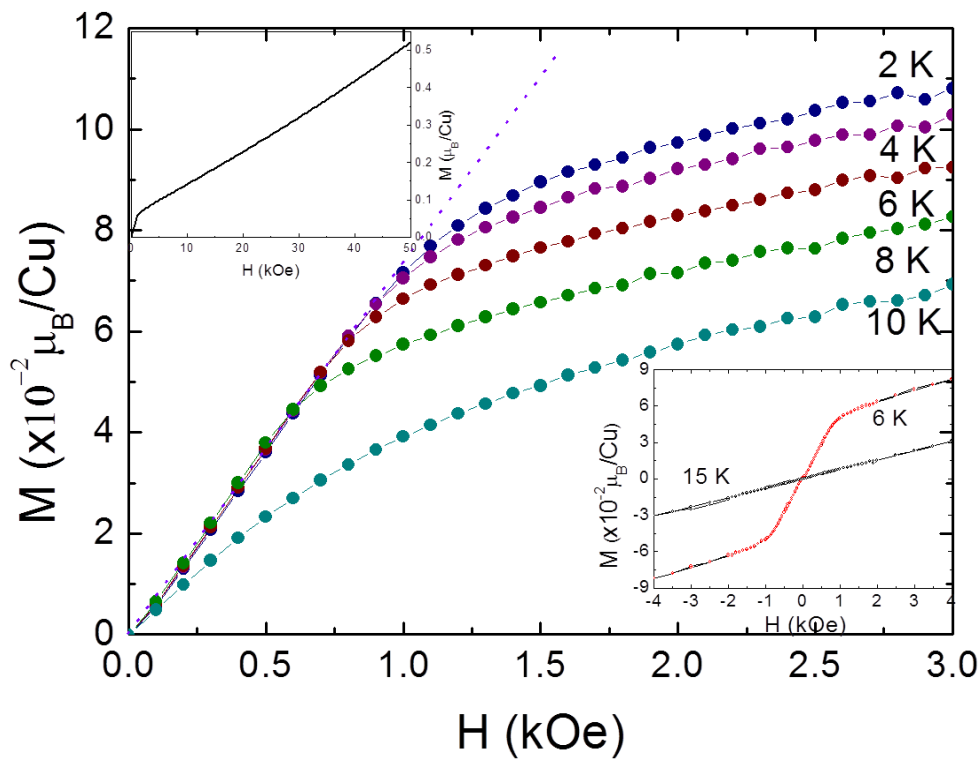

**Figure 4: Field dependence of magnetisation for post-pressurized $Cs_2CuCl_4$.** The field dependence of magnetization of the post-pressurized $Cs_2CuCl_4$ with each temperature from 2 K, 4 K, 6 K 8 K and 10 K is shown in the main figure. A possible transition magnetic field $H_c$ could be determined from the deviation of the linear line at low field crossover. As the temperature increases, $H_c$ is decreased and not observed above a transition temperature of 9 K. The magnetization versus field curves shows the S-shape with reversibility and linear behaviour below and above 9 K, respectively, in the lower inset. The upper inset shows the magnetization curve with linear behaviour at 2 K, from 1 kOe to 50 kOe.

While the magnetization versus field curve above the transition temperature of $T_s = 9$ K for post-pressurized $Cs_2CuCl_4$ is linear, indicating a paramagnetic phase, its curve below 9 K is a reversible *S*-shape, as shown in the bottom inset of figure 4. Figure 4 depicts the field dependence of magnetization at different temperatures below 10 K. Below $T_s = 9$ K the magnetization curves increase sharply with magnetic field with a universal gradient before slowly saturating above a magnetic field ($H_d$), which is determined from the point of deviation from the low field susceptibility indicated in the figure. The deviation magnetic field $H_d$ decreases as temperature increases up to 9 K and could not be observed above 10 K. The magnetic behaviour of the post-pressurized $Cs_2CuCl_4$ single crystal is similar to that seen in helimagnetic like magnetization curves [38,39], in particular the decrease and disappearance of the transition temperature $T_s$ with increasing field, and of the field $H_d$ with increasing temperature. The magnetisation curve above 1 kOe for post-pressurized version is linearly increased up to 50 kOe, as shown in the upper inset of figure 4. The lack of saturation in the M-H curves may suggest that the field-induced ferromagnetic-type order, shown in figure 3 (a), is based on the disordered antiferromagnetic phase.

The different magnetic properties of pre- and post-pressurized $Cs_2CuCl_4$ single crystals are likely to be caused by the change of antiferromagnetic spin exchange coupling between Cu ions, caused by structural distortion or structural phase transition. The crystal structure of the pre-pressurized $Cs_2CuCl_4$ is orthorhombic with lattice parameters of $a$ = 0.97644(2) nm, $b$ = 0.76143(3) nm, and $c$ = 1.23988(5) nm at room temperature as shown in figure 1(a), consistent with the reported results [15]. An X-ray diffraction study under high pressure for a $Cs_2CuCl_4$ polycrystalline sample reported that the volume and length of all crystallographic directions systematically decreases with increasing pressure up to 40 kbar and are reversible as pressure is released without any structural phase transition [40]. It was reported by P. T Cong et. al, [41] that, in the doped range of $1 \leq Br \leq 2$ in $Cs_2CuCl_{4-x}Br_x$, the magnetic behaviours are significantly different from spin liquid phase in $Cs_2CuCl_4$ due to change from O-type orthorhombic (Pnma) to T-type tetragonal (I4/mmm) structure.

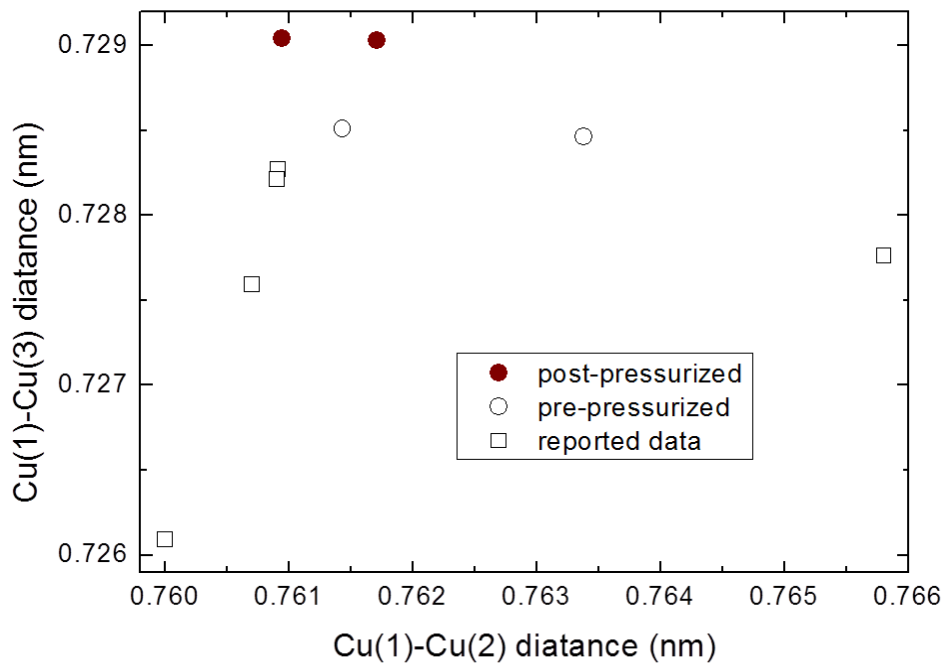

**Figure 5: Distances between Cu ions of $Cs_2CuCl_4$.** The distance between Cu(1) and Cu(3) (related to $J'$) for post-pressurized $Cs_2CuCl_4$ (solid symbols) is slightly longer than those for pre-pressurized and the reported samples (open symbols) [15, 40, 42-44], however, the other (related to $J$) is similar to them.

The lattice parameters of the post-pressurized $Cs_2CuCl_4$, with the values $a = 0.97663(4)$ nm, $b = 0.76094(3)$ nm, and $c = 1.24146(4)$ nm, are similar to those of pre-pressurized without any crystal structure changes, as shown in Table I. These cell parameters are consistent with the reported x-ray data [15, 40]. While the differences the lengths of *a*- and *b*- axis between the pre- and post-pressurized $Cs_2CuCl_4$ are an order of $10^{-2}$ %, the differences in the length of *c*-axis between them could not be ignored with an order of $10^{-1}$%. Using our x-ray experiment results, we plot the distances between Cu(1) and Cu(2) ions and between Cu(1) and Cu(3) ions for the pre- and post-pressurized $Cs_2CuCl_4$ with the reported resulted [15,40, 43-45], as shown in figure 5. While the distance between Cu(1) and Cu(2) of the post-pressurized samples is similar to the pre-pressurized samples and the reported results, the distance between Cu(1) and Cu(3) is slightly longer than them. Due to the slight change between distance of Cu ions, NNN interaction *J'* on the triangular lattice for the post-pressurized $Cs_2CuCl_4$ could be changed, relative to the pre-pressurized value (*J'* = 0.125 meV). As a result, it could cause a spin rearrangement on the triangular frustrated antiferromagnet system with a change in spin frustration ratio.

|  | Pre-pressurized $Cs_2CuCl_4$ (i) | Post-pressurized $Cs_2CuCl_4$ (ii) | Difference between (ii) and (i) |
|---|---|---|---|
| a (Å) | 9.7644(2) | 9.7663(4) | 0.0019 |
| b (Å) | 7.6143(3) | 7.6094(3) | -0.0049 |
| c (Å) | 12.3988(5) | 12.4146(5) | 0.0158 |
| V (Å$^3$) | 921.84(5) | 922.6(6) | 0.76 |
| Cu(1)-Cu(2) | 7.6143(3) | 7.6094(3) | -0.0049 |
| Cu(1)-Cu(3) | 7.2851(1) | 7.2804(6) | 0.0053 |

TABLE I: The comparisons for crystal structures between the pre- and post-pressurized $Cs_2CuCl_4$ single crystals at room temperature.

As a possible scenario for the interpretation of our results, the magnetic structure for post-pressurized $Cs_2CuCl_4$ would be considered to be a less frustrated and more quasi-one dimensional antiferromagnet spin-1/2 state with ordered magnetic phase. Theoretical studies suggested that the various ground magnetic phases in $Cs_2CuCl_4$ might exist as, for example, collinear antiferromagnet, spiral, and dimer phases, which are dependent of *J'* and second NN chain exchange coupling $J_2$ [25,28,29]. The magnetic phase of $Cs_2CuCl_4$ could be sensitive to changes in the interactions of a few percent in magnitude of the largest exchange constant [28] and ferromagnetic couplings in second NN chains are generated by fluctuations [25,28,29]. Tiny structural modification may induce a complete change of the ground state due to the high sensitivity of the exchange coupling constants in the triangular frustrated antiferromagnet spin-1/2 system $Cs_2CuCl_4$. We suggest that with pressure treatment the spin liquid gives way to new magnetic ground state with ferromagnetic-type interactions. In order to explain this phase in post-pressurized Cs$_2$CuCl$_4$, further theoretical as well as experimental study will be required.

**Acknowledgments**

We would like to thank G. G. Lonzarich, Victor Eremenko, Valentyna Sirenko, S. E. Rowley, P. Nahai Williamson, C. J. Pickard, Patricia L. Alireza, Swee K. Goh and S. E. Dutton for useful discussions. We acknowledge support from KAZATOMPROM Kazakhstan, EPSRC and Jesus College, Cambridge. The work at SNU was supported by National Creative Research Initiative (2010-0018300). Work at Rutgers is funded by the Gordon and Betty Moore Foundation's EPiQS Initiative through Grant GBMF4413 to the Rutgers Center for Emergent Materials. We thank CamCool Research Ltd. for providing the pressure cells used in this work.

**Additional information**

* Correspondence should be addressed to S. S. Saxena (**sss21@cam.ac.uk**), H-J. Kim (**hjk37@cam.ac.uk** ) and C. R. S. Haines (**crsh2@cam.ac.uk**)